# A Photonic Topological Mode Bound to a Vortex

Adrian J Menssen[1*], Jun Guan[2*†], David Felce[1†], Martin J Booth[2] and Ian A Walmsley[1,3]

[1]University of Oxford, United Kingdom, Department of Atomic and Laser Physics

[2]University of Oxford, United Kingdom, Department of Engineering Science

[3]Imperial College London, United Kingdom

(*)Correspondence to: adrian.menssen@physics.ox.ac.uk, jun.guan@eng.ox.ac.uk

†these authors contributed equally

**Topological photonics sheds light on some of the surprising phenomena seen in condensed matter physics that arise with the appearance of topological invariants[1,2]. Optical waveguides provide a well-controlled platform to investigate effects that relate to different topological phases of matter, providing insight into phenomena such as topological insulators and superconductors by direct simulation of the states that are protected by the topology of the system. Here, we observe a mode associated with a topological defect[3] in the bulk of a 2D photonic material by introducing a vortex distortion to an hexagonal lattice and analogous to graphene[4,5]. These observations are made possible by advances in our experimental methods. We were able to manufacture uniform large two-dimensional photonic crystal structures, containing thousands of identical waveguides arranged in two dimensions[21], and we developed a new method to excite multiples of these waveguides with a well-defined light field. This allows us to probe the detailed spatial features of topological defect modes for the first time. The observed modes lie mid-gap at zero energy and are closely related to Majorana bound states in superconducting vortices[3,6]. This is the first experimental demonstration of a mode that is a solution to the Dirac equation in the presence of a vortex, as proposed by Jackiw and Rossi[7,8].**

Phenomena associated with the topological characteristics of physical systems are of wide-reaching interest in many fields in physics, with applications ranging from condensed matter physics to particle physics[7] and cosmology[9]. Most prominently topology has been applied in condensed matter physics, where the importance of the topology of the band structure was first recognised in the discovery of the integer quantum Hall effect[10]. Subsequently many classes of topological insulators and superconductors have been discovered[3].

A key characteristic of a system with non-trivial topology is the presence of topological invariants, integer numbers that classify the topological structure. They are preserved under smooth, 'homotopic' deformations of the Hamiltonian. At the boundary between domains governed by different such invariants, where the topology abruptly changes, a topological defect occurs. Localised at these defects are states protected by the topology of the system: they are robust to errors in the underlying Hamiltonian. These 'edge states' have been investigated extensively in photonic platforms[11–15]. Their study generated important insight into the physics of topological insulators[13] and spawned technological advances such as the development of topological lasers[16], where lasing occurs in edge states, protected from imperfections.

Here, in contrast, we investigate for the first time a state bound to a vortex, a point defect, in the bulk of a 2D photonic material (photonic graphene). These modes are 'zero modes', as they lie mid-gap, at zero energy. The vortex-bound zero modes we consider here are of significant interest to solid-state physics. In topological superconductors they support Majorana bound states at the vortex core[3,6,17–19]. The vortex defect we realise here are photonic analogues of the vortices in topological superconductors explored by Reed, Green[18] and Volovik[17]. Majorana bound states at these vortices are a promising candidate for the realisation of a topological quantum computer[20].

The Jackiw-Rossi[7] model was originally introduced in a quantum field theory context to describe fermions in a 2+1D system coupled to a complex scalar field in the Dirac equation. Jackiw and Rossi[7] demonstrated that if the scalar field contains a vortex it supports localised zero-energy solutions at the vortex core. Witten[9] also considered the Jackiw-Rossi type vortex defect in the context of superconducting cosmic strings.

Particles in a tight binding model of graphene can be shown to obey a Dirac equation near the 'Dirac' cones in the band structure. Hou, Chamon and Mudry[4] predicted zero modes in graphene which realise the Jackiw-Rossi model[8]: Small perturbations to the graphene lattice induce the scalar-field coupling to the effective Dirac equation governing the graphene tight binding model.

A photonic platform allows for a high degree of control over the system parameters and is thus a powerful tool to investigate topological effects, where in solid state systems this degree of control is notoriously difficult to engineer. The vortex topological defect is realised by introducing a distortion to the waveguide position in a hexagonal waveguide lattice as was suggested by Iadecola et. al.[5]. We show adiabatic transport of the zero mode as the vortex is moved, as well as topological protection against imperfections of the waveguide lattice. Our research promises to open new avenues in topological photonics[22], and provide new insights in the physics of topological solid state systems.

The wavefunction evolution generated by a tight binding Hamiltonian can be mapped directly to the coupled mode equations of light propagating through a photonic crystal, where the time dimension in the Schrödinger equation takes the role of the propagation direction of the light through the crystal:

$$i\frac{\partial \psi_{\boldsymbol{r}}(z)}{\partial z} = \sum_{j=1}^{3} H_{\boldsymbol{r},\boldsymbol{r}\pm\boldsymbol{s}_j}\psi_{\boldsymbol{r}\pm\boldsymbol{s}_j}(\mathrm{z}) \qquad (1)$$

$\psi_{\boldsymbol{r}}(z)$ is the electric field amplitude at lattice site $\boldsymbol{r}$ and z is the length of the crystal the light has propagated through, $\boldsymbol{r}$ is the coordinate of a site on the hexagonal lattice and $\boldsymbol{s_j}$ points to its $\boldsymbol{j}$th nearest neighbour (see Fig. 1).

The complex scalar field $\Delta(\boldsymbol{r})$ containing the vortex can be simulated by introducing a distortion $\delta t_{\boldsymbol{r},\boldsymbol{s}_j}$ to the nearest neighbour hopping strength $t$ of graphene[4]:

$$\begin{aligned} H_{\boldsymbol{r},\boldsymbol{r}\pm\boldsymbol{s}_j} &= -t - \delta t_{\boldsymbol{r},\boldsymbol{s}_j} \\ \delta t_{\boldsymbol{r},\boldsymbol{s}_j} &= \frac{1}{2}\Delta(\boldsymbol{r}) e^{i\boldsymbol{K}_+ \boldsymbol{s}_j} e^{2i\boldsymbol{K}_+ \boldsymbol{r}} + c.c., \end{aligned} \qquad (2)$$

$\boldsymbol{K}_+ = \left(\frac{4\pi}{3\sqrt{3}a}, 0\right)$ marks the position of Dirac point in the reciprocal lattice. $a$ is the lattice constant.

The explicit form of $\Delta(\boldsymbol{r})$ containing a vortex at $\boldsymbol{r} = \boldsymbol{0}$ is:

$$\begin{aligned} \Delta(\boldsymbol{r}) &= \Delta_0(r) e^{i(\alpha + N\varphi)} \\ \Delta_0(r) &= \Delta_0 \tanh(r/l_0), \end{aligned} \qquad (3)$$

where $\varphi$ is the polar angle of $\boldsymbol{r}$, $l_0$ the width of the vortex, $\alpha$ the phase of the vortex and $N$ the vorticity, the topological invariant of the system. The sign of the vorticity determines which of the two triangular sub lattices of graphene (see Fig.1) supports the mode $(1 \text{ for sublattice A}, -1 \text{ for}$ sublattice B). In this work we set its value to 1: the mode is confined to sub lattice A, whereas the distortion is applied to sub lattice B.

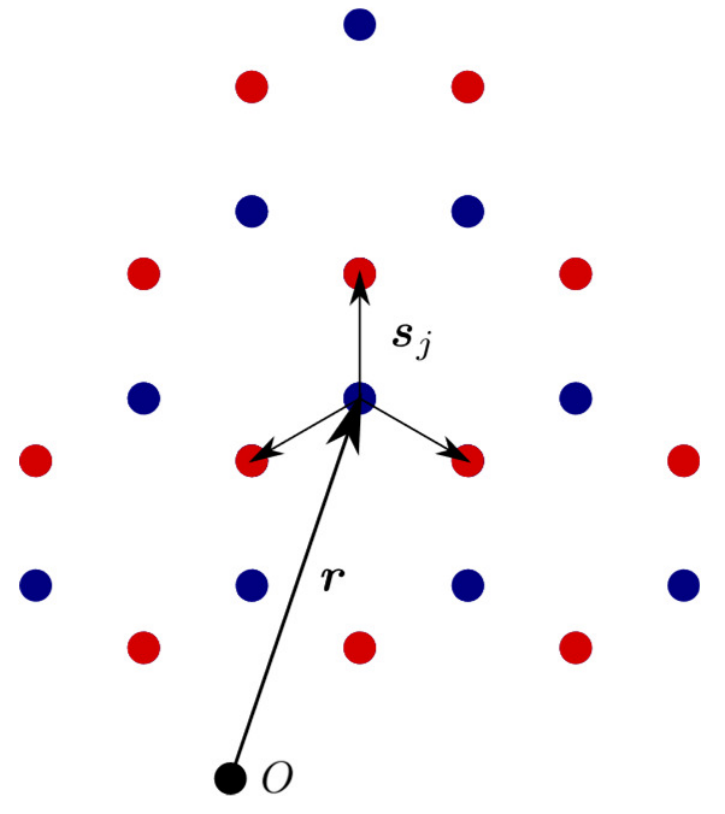

*Fig. 1* Hexagonal lattice with two triangular sublattices (A and B) in blue and red, lattice vector $\boldsymbol{r}$ and vectors between nearest neighbours $\boldsymbol{s}_j$ indicated.

The spatial dependence of the zero mode-wavefunction $\phi_A(r)$ localised on sub lattice A (a solution to the mode-coupling equations (1)) is given by[4,7,18]:

$$\phi_A(r) \sim e^{i(\frac{\alpha}{2}+\frac{\pi}{4})} e^{-\int_0^r dr' \Delta_0(r')}. \qquad (4)$$

For photons in a lattice, the wavefunction is modulated by $e^{i\boldsymbol{K}_+ \boldsymbol{r}}$ giving the expression for the electrical field strength as a function of the lattice vector[5]:

$$E(\boldsymbol{r}, z, t) \sim \mathrm{Re}[(\phi_A(r) e^{i\boldsymbol{K}_+ \boldsymbol{r}} + c.c.) e^{i k_\omega z - i\omega t}], \qquad (5)$$

where $\boldsymbol{r}$ points to a lattice site on the supporting sublattice A.

The vortex in $\Delta(\boldsymbol{r})$ and the zero-mode state bound to it are visualised in Fig. 2 a). The distortions $\delta t_{\boldsymbol{r},\boldsymbol{s}_j}$ of the coupling Hamiltonian are implemented with small shifts in the wave-guides' positions (illustrated in Fig. 2 b)), assuming exponential decay of the field away from the waveguides. The exponential decay was measured experimentally for different waveguide positions and relative orientations (see Appendix D). These data allow us to implement the hopping matrix-elements (Eq.2) faithfully. We note that it is the collective effect of small distortions applied to every lattice site which gives rise to the topologically confined mode at the centre of the vortex defect.

Our photonic crystal lattices are fabricated in glass through femtosecond laser direct writing[21,23] (see also Appendix C)). To excite the bulk zero mode, we developed a method based on an SLM to simultaneously illuminate multiple waveguides with beams of individually controlled phase and amplitude, as illustrated in Figure 2 d). Multiple beams (up to 13 in the present work) are directed at individual waveguides near the centre of the vortex. Phases and amplitudes of all beams are tuned to excite the zero mode (setup described in Appendix A). We use a gradient descent algorithm to optimise the phases and amplitudes for maximum overlap with the zero mode (see Appendix B). We note that to observe the spatial features of bulk zero mode, such as its localisation on one triangular sublattice, exciting the mode with a single waveguide is insufficient. Each waveguide lattice consists of 1192 waveguides that are arranged to form a hexagonal lattice. The distance between lattice sites is around 10 μm. This is large enough to ensure suppression of next nearest neighbour coupling. We estimate the ratio of nearest to next nearest neighbour coupling strength to be less than 5%. The dimension of the waveguide lattice is around 400×400 μm. First, we demonstrate a stationary photonic bulk zero-mode. The vortex distortion is located at the centre of the lattice, as shown in Fig. 2a/b). In Fig. 2c), comparing the experimental result (top) and the theoretical calculation (bottom), we see that they match very well. The brightest peaks are in each case the central one and six weaker peaks that form a large hexagon around the central mode. The waveguide mode intensity pattern is governed by equations (4) and (5). For the vortex phase we chose $\alpha = -\frac{\pi}{2}$, which results in the characteristic hexagonal pattern we observe. The modulation of the wavefunction by $e^{i\boldsymbol{K}_+ \boldsymbol{r}}$ (resulting from the linearization around the Dirac points) is offset by the phase of the vortex $\alpha$ such that the centre waveguide and the six next nearest neighbours on sublattice A carry most of the mode intensity. The light in the zero mode is tightly confined to the centre of the vortex and decays quickly outside the radius of the vortex $l_0 = 20$ μm, with a decay length governed by equation (4). Most of the intensity is confined to the sublattice which carries the bulk zero-mode, as designed. To quantify the degree to which this intensity pattern represents the zero-mode, we introduce the ratio of light intensity between the two sublattices

$$\gamma_{AB} = \frac{\text{Light intensity in sublattice A}}{\text{Light intensity in sublattice B}}$$

as a measure of fidelity for the excitation of the mode, which should be confined to one sublattice only. For a mode confined to sublattice A, we expect $\gamma_{AB} \gg 1$. The measured mode displayed in the top panel of Fig. 2 c) has $\gamma_{AB} = 5.9$.

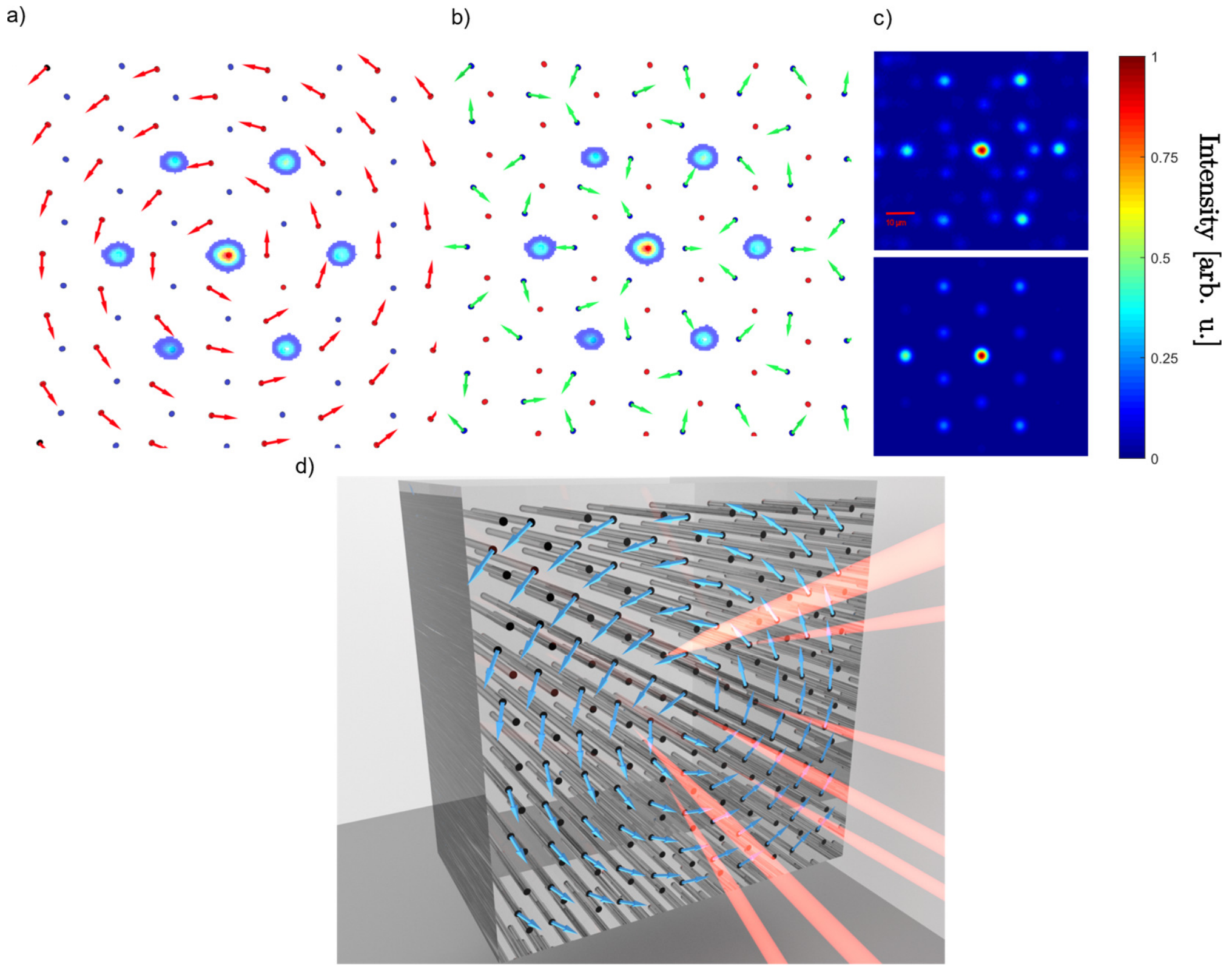

*Fig. 2* a) Photonic lattice with $\Delta(\boldsymbol{r})$ represented by red arrows. Bulk zero-mode intensity for dominant modes indicated at the centre of the vortex. For visualising the order parameter, the complex function $\Delta(\boldsymbol{r})$ is mapped onto a vector field through the correspondence $\mathbb{C} \rightarrow (Re[\Delta(\boldsymbol{r})], Im[\Delta(\boldsymbol{r})])$, we display the vectors for each point $\boldsymbol{r}$ on sublattice B. b) Shift of the waveguide positions from the graphene lattice configuration, which implements the topological vortex distortion. The direction of the shift is indicated by green arrows. c) Comparison between the camera image of an observed stationary topological bulk zero mode (Top) and its theoretical simulation (Bottom). The theory has been convolved a with a Gaussian to approximate the modes of the waveguides. The ratio of light intensity in the centre waveguide to the averaged intensity in a waveguide lying on the outer bright hexagon is approximately 3:1, which is consistent with the expected mode decay of the vortex-bound wavefunction (Equation (5)). d) Illustration of multi-beam excitation of the photonic lattice. An array of phase and amplitude tuned beams directed at individual waveguides carrying the majority of the intensity is used to achieve coherent excitation of the zero mode in the centre of the vortex.

Next, we translate the vortex zero mode transversely across the waveguide lattice by adiabatically shifting the vortex from one side of the lattice to the other by around 100 µm. To ensure adiabaticity we need at least 4 cm of propagation distance, this was determined by propagating the mode numerically using the model Hamiltonian obtained from the prior measurement of the coupling coefficients (see Appendix D). In the experiment, a 9 cm long chip is used. We can observe transverse translation of the zero-mode with most of the transmitted intensity confined around the centre of the shifted vortex and in the correct sublattice (with a ratio of $\gamma_{AB} = 4.7$, Fig. 3 a), top panel). The measured mode is no longer symmetrical, due to small variations in waveguide lattice fabrication which can easily shift the relative intensities within waveguides in zero mode. In the bottom panel of Fig. 3 a), we show what happens if we try to excite the mode at a position in the lattice where there is no vortex present. As expected we are not able to excite a mode and see no light transport. The ratio of intensities $\gamma_{AB} = 0.67$ is $\ll 1$.

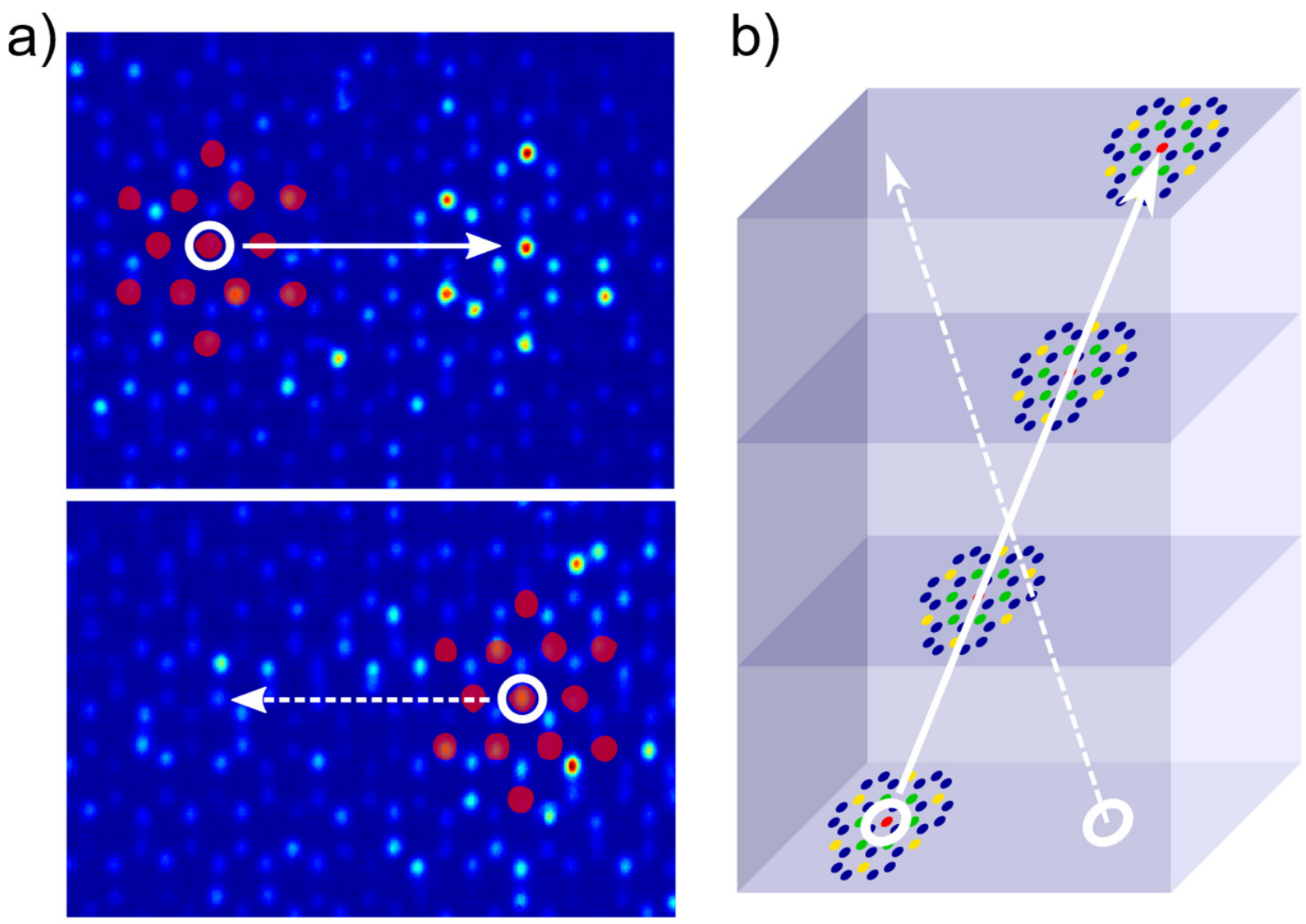

*Fig. 3 a)* Top panel: Recorded bulk zero-mode transverse translation, in which the zero-mode is translated from left (at input side schematically shown in b)) to right (at the output side) of the lattice. Red dots indicate the position of the 13 beams used to excite the zero mode at the input. The circled waveguide marks the position of the centre at the input side. $\gamma_{AB} = 4.7$. Bottom panel: the same number of waveguides are excited on the input side of the lattice, where no vortex is present. At output side, $\gamma_{AB} = 0.67$ also, the total light intensity in the displayed region is substantially less as most of the light is scattered into the rest of the lattice. The colour scale is normalised to the brightest peak in both pictures. b) Illustration of mode shifting by adiabatically translating (solid arrow) the centre of the vortex (dotted pattern) along the longitudinal direction of waveguide lattice; the dashed arrow indicates the direction the vortex moves in.

To demonstrate that the zero-mode is topologically protected against random errors of the lattice, we deliberately introduce an error to the position of the waveguides by shifting them by a random distance, sampled from a two-dimensional uniform distribution, where the radius $r_d$ of the distribution corresponds to the maximum shift applied. We note that our mode is protected against any error which preserves chiral symmetry, such as random errors in wave guide position. In Fig. 4, we show the recoded zero-modes at four different distortion levels with $r_d = \{0,200,400,600\}$ nm respectively. The systematic topological distortion, which is introduced to the hexagonal lattice for the formation of the vortex, is approximately 850 nm. We can observe that the zero-mode remains visible even when large random errors in waveguide position are introduced. The larger the error distortion, the more light leaks into the other sublattice. We measure a steadily decreasing amount of light in the correct sublattice $\gamma_{AB} = \{3.8, 2.9, 2.2, 2.1\}$ as the error distortion increases.

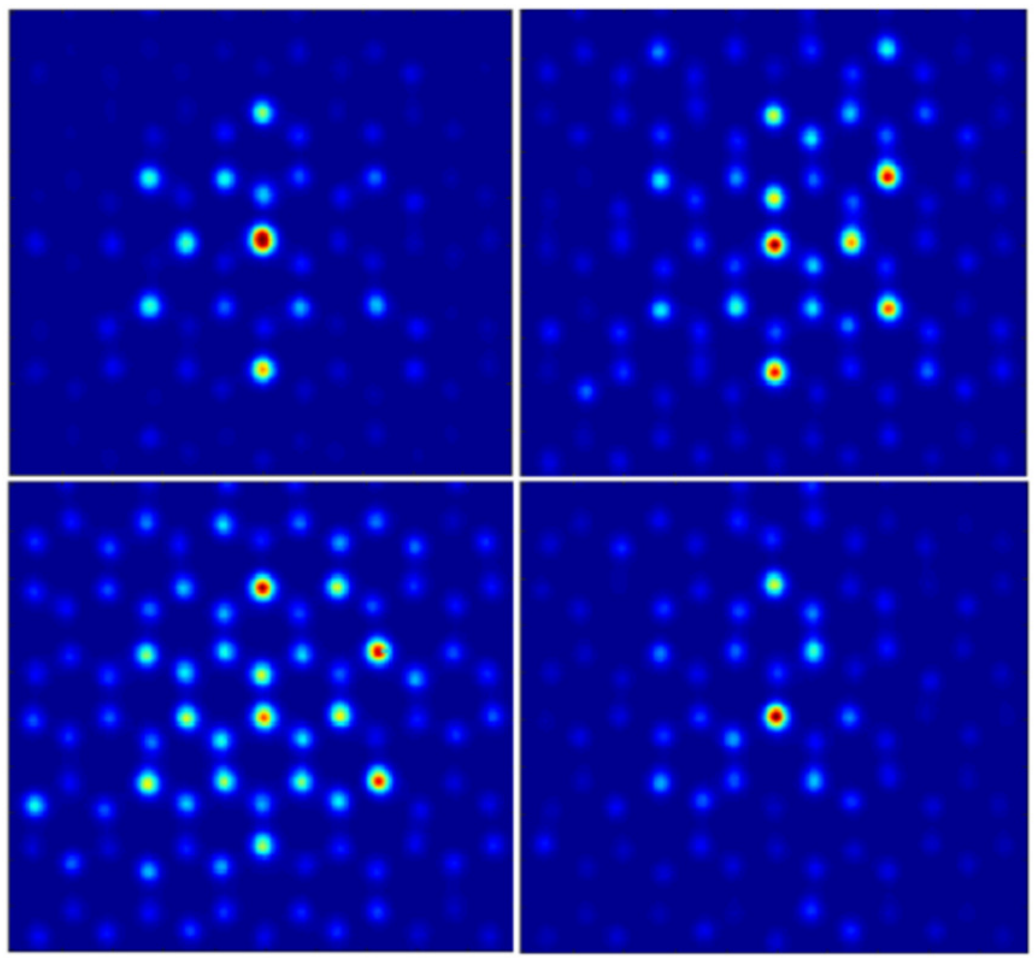

*Fig. 4* Topological mode with random error distortions. Top left: no error, Top right: maximum 200 nm, Bottom left: maximum 400 nm, Bottom right: maximum 600 nm. The ratios of light intensity in the zero-mode sublattice to the non-mode supporting one are $\gamma_{AB} =$ 3.8, 2.9, 2.2, 2.1 respectively. The colour map of each image is normalised independently.

We have demonstrated for the first time an experimental implementation of the Jackiw-Rossi model and observed the characteristic topological vortex zero mode in photonic lattice. We demonstrate that the mode can be adiabatically shifted across photonic lattice and that it is topologically protected. The degree of control we have demonstrated over these localised, protected topological bulk zero-modes may in the future enable applications such as the protection of information encoded in quantum states[24,25] against inevitable fabrication errors in linear optical circuits by injecting multiple entangled particles into different protected vortex modes. Moreover this demonstration also provides the possibility for further studying models employed in high energy physics, as well as new implementations of topological lasers other than those which have already been demonstrated in edge-states[16,26].

**Acknowledgements:** We especially want to thank for their constructive input Claudio Chamon, Tom Iadecola and Thomas Schuster. We thank Bryn Bell for sharing his knowledge in topological photonics. Furthermore, we would like to acknowledge Steve Simon and Glenn Wagner for providing valuable insights into solid state theory, Helen Chrzanowski for her experimental expertise and Steve Kolthammer for initial discussions. We thank Patrick Salter for sharing his expertise in aberration correction techniques. AJM is supported by the Buckee Scholarship at Merton College. This work was funded by the EPSRC through the Programme Grant EP/K034480/1, and by the European Union through the ERC Advanced Grant MOQUACINO.

Author contributions: AJM conducted the theoretical simulations, conceived the idea for the SLM based method of exciting multiple waveguides and wrote the paper with input from the other authors. JG built the adaptive femtosecond laser writing system, implemented the adaptive aberration correction and fabricated the chips. AJM and DF built and conducted the experiment. DF worked on the development and implementation of the method to control phase and amplitude of the light field. IAW, AJM, JG and MJB conceived the project. IAW and MJB supervised the project.

# A Photonic Topological Mode Bound to a Vortex

Adrian J Menssen[1*], Jun Guan[2*†], David Felce[1†], Martin J Booth[2] and Ian A Walmsley[1,3]

**Affiliations:**

[1]University of Oxford, United Kingdom, Department of Atomic and Laser Physics

[2]University of Oxford, United Kingdom, Department of Engineering Science

[3]Imperial College London, United Kingdom

(*)Correspondence to: adrian.menssen@physics.ox.ac.uk, jun.guan@eng.ox.ac.uk

†these authors contributed equally

## A – Experimental setup

To excite the desired topological mode in the photonic lattice we need phase-stable illumination of many waveguides with individual amplitude and phase control. The setup is shown in Fig. S1. A Spatial Light Modulator (SLM) provides pixel-wise control of the light field[27–32] in terms of phase and amplitude. The SLM in this setup is a reflective Ferroelectric Liquid Crystal device (FLC-SLM). Each pixel effectively acts as a half-waveplate, where the angle of the optical axis can be continuously rotated by changing the voltage applied to each pixel. The SLM contains an array of 512×512 pixels. It is illuminated by a coherent CW-laser beam at 785 nm wavelength. Placing a polarising beam splitter after the SLM affords amplitude control of the light by rejecting its vertically polarised component. To achieve both amplitude and phase control, we adapt a super pixel method employed by Goordon et.al.[33], taking advantage of the continuous control of the light amplitude afforded by the FLC-SLM. Field contributions from neighbouring pixels are combined by filtering high spatial frequency components of the light field, forming a 2×2 "super pixel". Controlling the light amplitudes of four pixels provides enough degrees of freedom to control both, amplitude and phase of the light field. The low pass spatial filter is implemented by placing a pinhole in the Fourier plane of a lens after the SLM. We image the surface of the SLM onto the end-facet of the waveguide lattice chip.

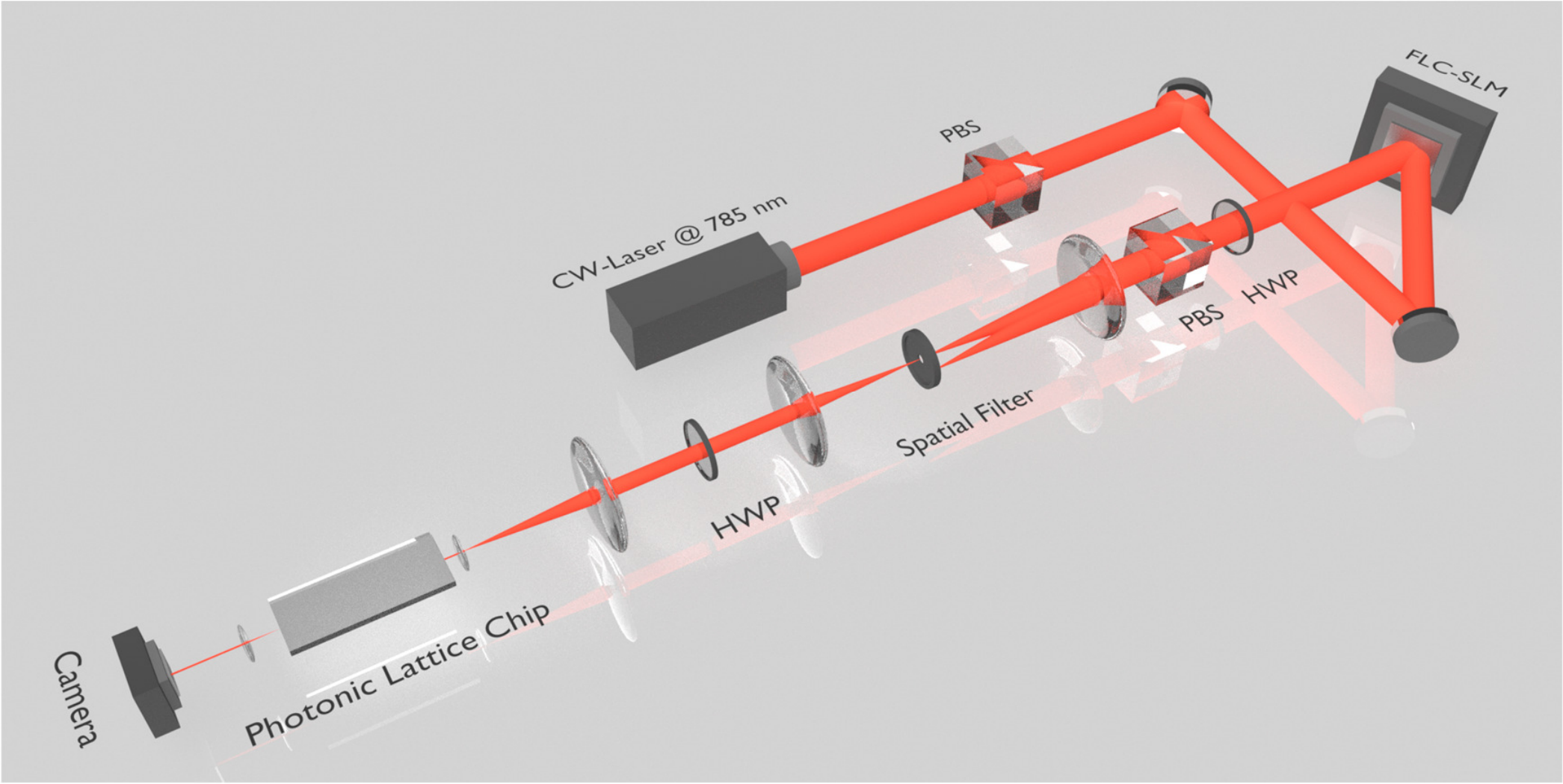

*Fig. S1* Experimental Setup: The coherent light source is a CW laser diode at centre wavelength of 785 nm. 1. We send the light through a Polarising Beam Splitter (PBS) to prepare the light in horizontal polarisation. 2. The beam is reflected from the surface of the Ferroelectric Liquid Crystal Spatial Light Modulator (FLC-SLM); each pixel rotates the light's polarisation state according to its optical axis orientation. 3. After passing through a Half Waveplate (HWP), the vertically polarised component of the light is rejected, implementing the desired field amplitude. 4. The light passes through a 4F-system with an off-axis pinhole, enabling phase control. 5. The polarisation of the light is prepared by a half-wave-plate (HWP). 6. The light beam is imaged onto the input -facet of the chip. 7. The light at the output-facet is imaged onto a camera.

We now illustrate the method light control in detail: An elementary result from Fourier optics is that a displacement in the Fourier plane corresponds to a linear phase gradient in the image plane. For example, in the one-dimensional case, if one obtains the image plane function $f$ by taking the inverse Fourier transform of a function $g$ in the Fourier plane, where $g$ has been shifted by a wave-vector $k_0$, then the result is the inverse transform of the unshifted $g$ with a phase gradient of $k_0$:

$$f(x) = \int\ dk\ g(k + k_0)e^{-ikx} = e^{ik_0x} \int\ dk'\ g(k')e^{-ik'x} \tag{S1}$$

Here the second equality follow by the substitution $k' = k + k_0$. This result allows us to assign different phases to different pixels. We place an iris in the Fourier plane with a displacement from the central axis chosen to give the desired phase gradient (Fig. S2) in the image plane. We choose the aperture of the iris so that the diffraction-limited spot size of the imaging system is increased to the size of a 2×2 group of pixels. For our 2×2 superpixel realisation, the upper limit for the aperture radius is $r_{max} = \lambda f/2d$, where $f$ is the focal length of the first lens in the 4F system and $d$ is the distance between pixels. Thus, the pixels are "blurred" into 2×2 "superpixels". The complex amplitude of the electric field vector from any one of these superpixels is then the sum of the complex amplitudes of the individual pixels. Each superpixel contains four pixels with complex amplitude vectors along each of the cardinal directions in the complex plane. We can therefore create any electric field amplitude within a circle in the complex plane.

| | | | |
|---|---|---|---|
| **0** | **π/2** | **π** | **3π/2** |
| **π** | **3π/2** | **0** | **π/2** |
| **0** | **π/2** | **π** | **3π/2** |
| **π** | **3π/2** | **0** | **π/2** |

*Fig. S2* The pixel phases generated by the phase gradient applied by the off-axis Fourier filter. The thicker borders denote the boundaries of 2×2 superpixels.

The final part of the input light preparation is another 4F telescope with an aspheric objective lens to achieve the required magnification, after which the face of the SLM is imaged onto the input facet of the lattice chip. We image the output of the chip using another aspheric lens and a camera. The output of the camera is fed back to the control PC. In the experiment, the information from the camera is used in real time to create a feedback loop to control the SLM.

We created a Graphical User Interface (GUI) which allows the user to define an array of beams, specifying the position at which beams are generated on the surface of the SLM, (and hence where they arrive in the image plane), their width, direction (or phase gradient, $k_x$ and $k_y$), intensity, and relative phase to other beams. This gives us a capability equivalent to the use of a pair of mirrors and a phase shifter for each of up to 400 beams, something which would clearly be impossible with bulk optics.

We implement the 2×2 superpixel method in our control software in the following way. First, we calculate a complex electric field array over the surface of the SLM from the beam information provided by the user in the GUI. This array has dimensions of half that of the SLM pixel array (in our case 256×256), so that the electric field has a value at the position of each 2×2 superpixel. As previously described, each pixel within the superpixel generates an electric field which points in a cardinal direction in the complex plane. To assign the illumination value of each pixel we calculate the projection of the complex electric field vector onto the cardinal vector of the superpixel (if this projection is negative then the pixel is set to zero). The resulting array is then fed to the SLM.

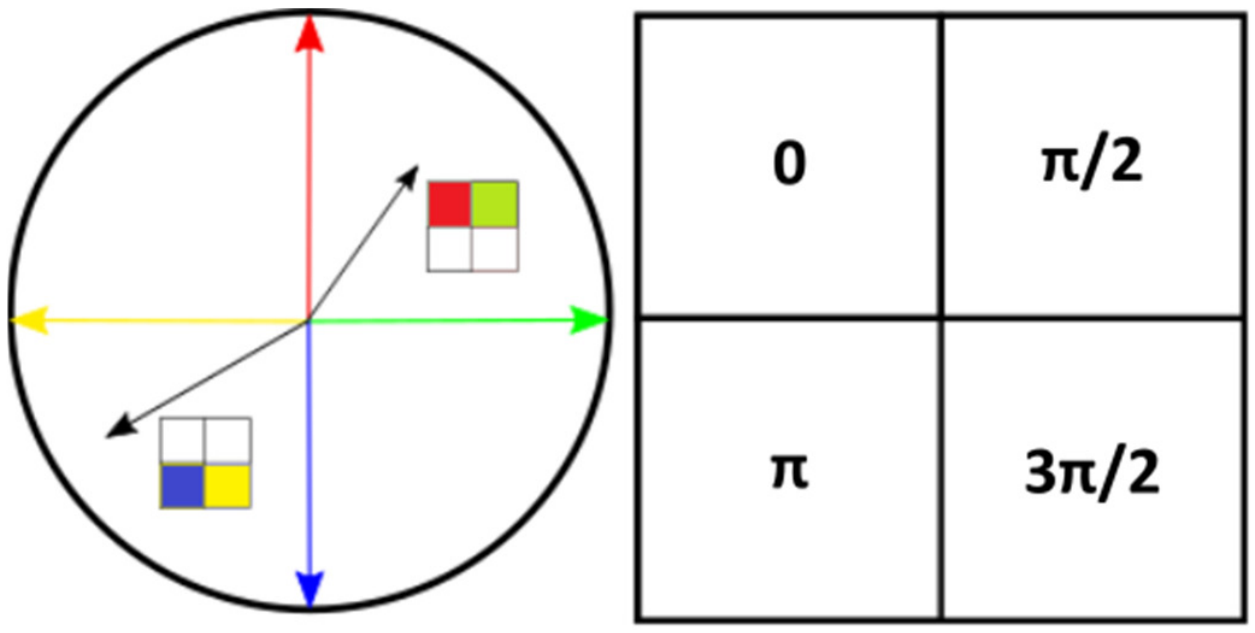

*Fig. S3* The illumination of the pixels is determined by the projection of the desired electric field vector onto the pixel's field vector. The amplitude of the superpixel is the sum of the individual pixel amplitudes and can be any value within a circle in the complex plane. The maximum amplitude possible is equivalent to that generated by ¼ of the pixels illuminated.

## B – Experimental procedure

To determine the waveguide positions at the input surface of the lattice chip, we scan a small beam across the input facet and record the total output intensity. At positions where the beam hits a waveguide there will be a high recorded intensity. The intensity at all points of the scan is plotted in a colour-map figure, shown in Fig. 4 a).

The waveguides which carry the intensity of the vortex zero mode are then selected by identifying them on the SLM scan colour-plot. Finally, we optimise the coupling into each waveguide. This is achieved by following a gradient descent method, allowing the position and direction of the beams to vary. This is analogous to using a pair of mirrors to walk and steer the beam but is achieved significantly faster and in a fully automated and reliable way.

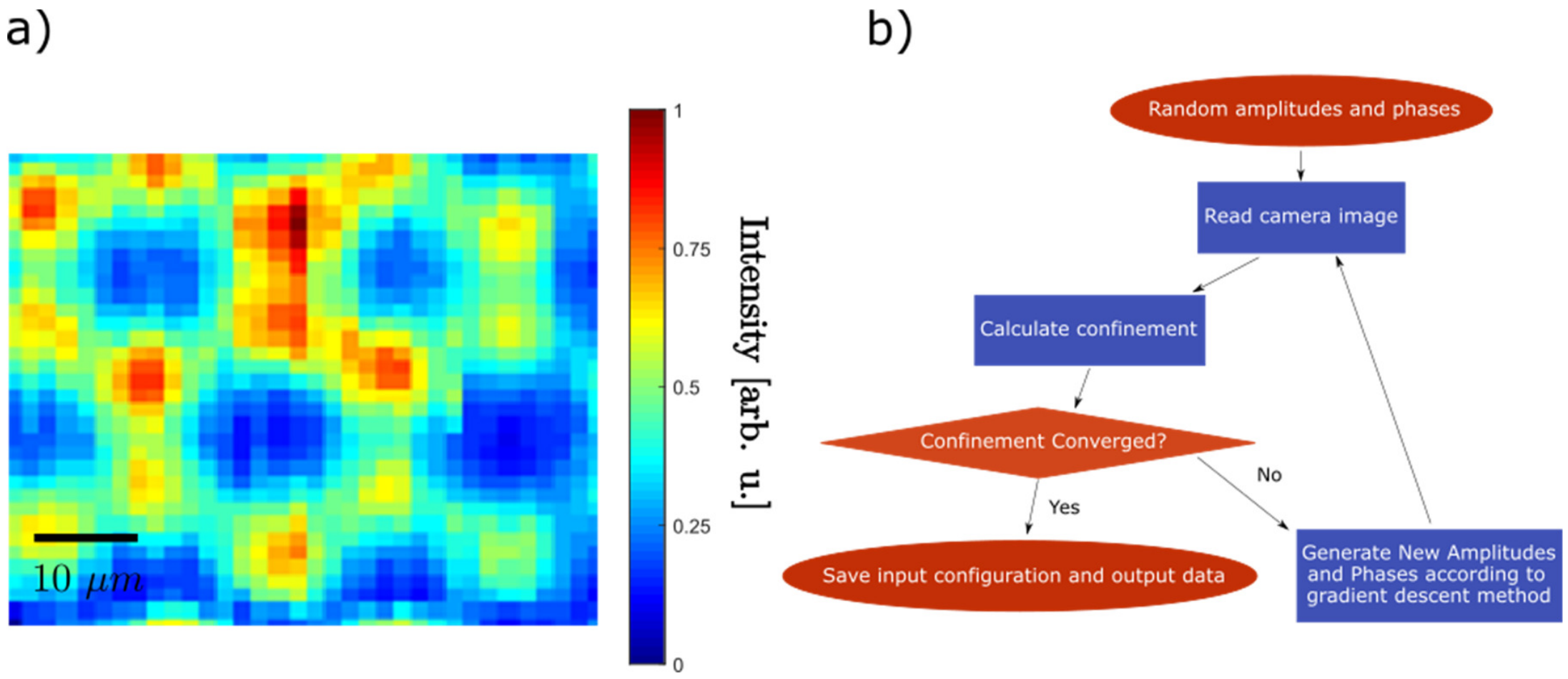

*Fig. S4* a) Total output light intensity plotted as a function of the position of a single x-y scanned input beam. This scan shows the location of waveguides at the input facet. b) Procedure of the optimisation routine maximising the coupling to the vortex zero mode.

Once beams have been coupled into every waveguide, we proceed to excite the mode. To do this we must find the correct relative intensity and phase to be coupled into each waveguide. These intensities and phases are predicted by the theory, but in our physical system each waveguide has a random input phase and there are imperfections in the lattice which mean that the optimum values of these parameters will be different from the values from the theoretical result. We employ a gradient descent method to find the optimum intensity and phase for each waveguide with a procedure illustrated in Fig. S4 b). The function we maximise is the ratio of output intensity confined within the region of the mode to the total intensity; in this way we do not optimise on the precise shape of the mode, so that when we observe the predicted shape at the output we can verify that we do detect the vortex zero mode. Once we achieve the optimum value of this ratio, the input configuration and output camera image are recorded. An optimisation result is shown in Fig. S5. For the adiabatically shifted mode we use 13 input beams, marked in locations in Fig. S5 a). The gradient descent algorithm is started with a set of 13 random phases and amplitudes. The value to be minimised is the ratio of the light intensity spread over the entire lattice and the light intensity within the region, where the zero mode is expected (green circle in Fig. S5 a/b). Once the mode is found, the procedure is terminated. In Fig. S5 c) the phase of one of the 13 excitation beams is plotted vs. the value of the optimisation target. Each point corresponds to a trial. In the present example it took ~150 trials to excite the zero mode successfully. We can clearly see that the mode is found at significantly lower value of the optimisation function. The apparent optimisation value threshold is marked with the red line in Fig. S5 c.

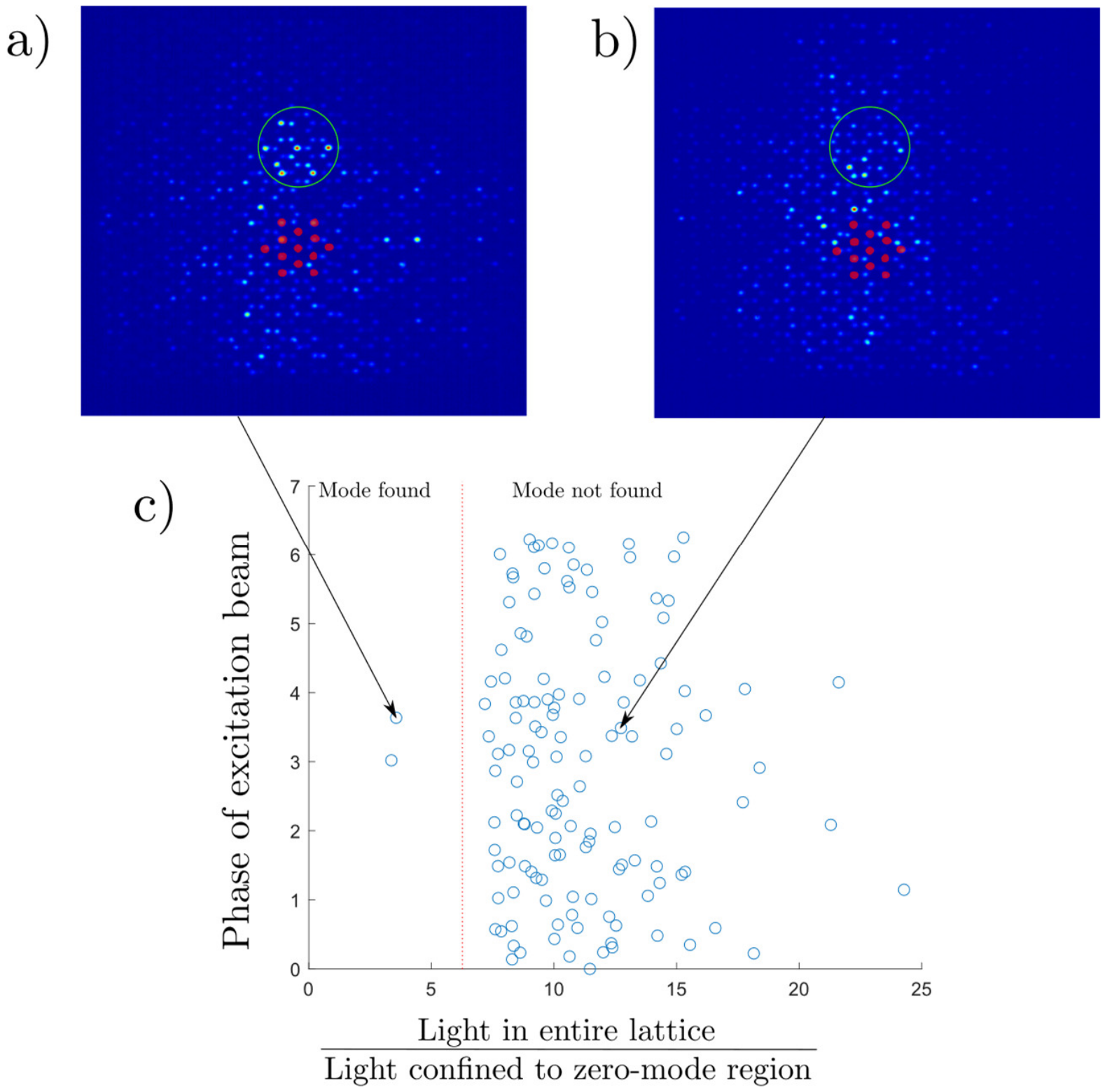

*Fig. S5* a) Successful excitation of the zero mode. Excitation beam pattern and location are indicated by 13 red dots. Green circle marks region of confinement for which the light intensity is maximised. The characteristic pattern of the zero mode is visible within the green circle: the value of the optimisation function is below the critical threshold. b) Unsuccessful excitation of the zero mode: the value of the optimisation function is above the critical threshold; no mode structure is visible. c) Phase of one of the exciting beams vs. the ratio of light intensity in the entire lattice and confined within zero-mode region. Points correspond to trials with varying random initial values for the phases and amplitudes passed to the gradient descent algorithm. A threshold in the optimisation value, which separates successful and unsuccessful excitation of the zero mode, is indicated with a red dotted line.

## C – Adaptive femtosecond laser writing

The laser that was employed to fabricate the waveguide lattices was the second harmonic of a regenerative amplified Yb:KGW laser (Light Conversion Pharos SP-06-1000-pp) with 1 MHz repetition rate, 514 nm wavelength, 170 fs pulse duration. The laser power beam was regulated through combination of a motorised rotating half waveplate and a polarization beam splitter before being phase-modulated by a liquid-crystal on silicon spatial light modulator (X10468-09(X), HAMAMATSU PHOTONICS K.K). Then the modulated beam was focused inside a glass chip with a 0.5 NA objective lens. The glass chip, which was fixed on a three-axis air bearing stage (AerotechABL10100L/ABL10100L/ANT95-3-V), was transversely scanned relative to the focus to inscribe waveguide. The waveguide lattices were written in borosilicate glass (Corning EAGLE 2000), with scan speed of 15 mm/s and pulse energy of 90 nJ at sample surface. The lattices were written inside the glass chips between 650 μm and 50 μm beneath the top surface.

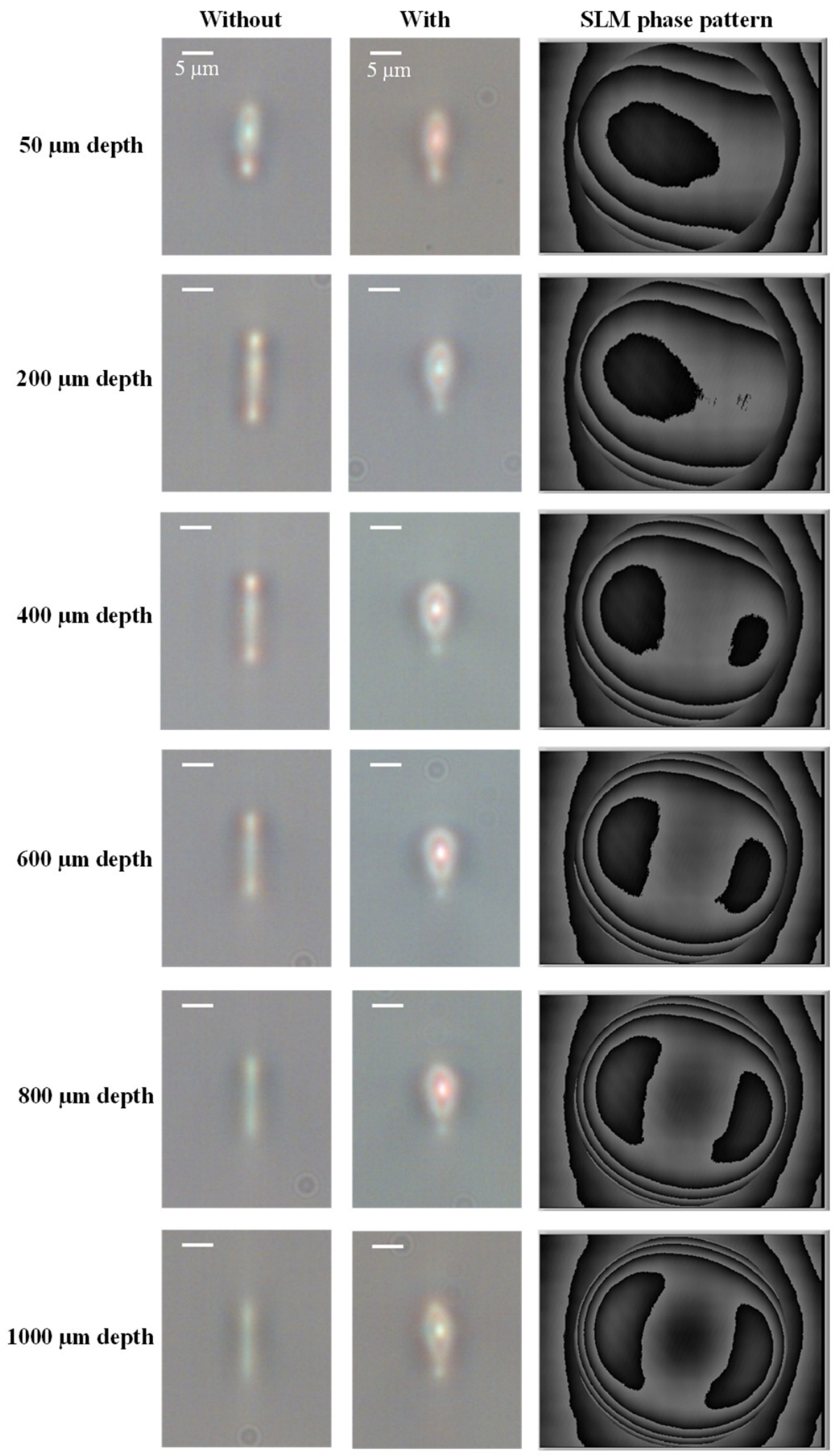

*Fig. S6* Comparison between single waveguides without and with aberration correction from depth 50 µm to 1000 µm; the first (from left to right) two columns are waveguide fabricated without and with correction of part II aberration; part I aberration had be corrected for all of them; third column are the SLM phase pattern used for correction of part I and II aberrations at corresponding depth; all waveguides were fabrication with the same scan speed of 15 mm/s and pulse energy of 90 nJ.

To fabricate uniform waveguides across the depth, we implemented automatic depth-adaptive aberration correction during fabrication. The aberration at the writing depth consisted of two parts: part I is the aberration introduced by the optics of the laser writing system; part II is the aberration caused by the refractive index mismatching between the objective immersion medium (air here) and the glass chip, which is depth-dependent. Part I aberration can be corrected experimentally by generating a fixed phase pattern on the SLM. Part II aberration can be corrected with phase pattern on SLM, which is generated based on the depth-dependent spherical aberration. The depth-dependent spherical aberration can be expressed as following equation (S2)[34] at the pupil plane of the objective lens.

$$\phi(r) = \frac{-2\pi d_{nom}}{\lambda}\left(\sqrt{n_c^2-(NAr)^2}-\sqrt{1-(NAr)^2}\right) \quad \text{(S2)}$$

where λ is the laser wavelength, $r$ is the normalised pupil radius, $n_c$ is the refractive index of the glass chip, the refractive index of air is 1, $d_{nom}$ is the nominal depth in the glass chip and $NA$ is the numerical aperture of the objective lens. The aberration contains a defocus element that is equivalent to a translation of the sample. To reduce the magnitude of the required SLM aberration correction, the defocus element was removed. The defocus-free spherical aberration can be written as:

$$\phi(r) = \frac{-2\pi d_{nom}}{s\lambda}\left(\sqrt{n_c^2 - (NAr)^2} - s\sqrt{1-(NAr)^2}\right) \quad \text{(S3)}$$
$$\frac{1}{s} = 1 + \frac{\langle \phi(r)', D_{n2}' \rangle}{\langle D_{n2}', D_{n2}' \rangle}$$
$$D_{n2} = \frac{-2\pi d_{nom}}{s\lambda}\sqrt{n_c^2 - (NAr)^2}$$

$\phi(r)'$ and $D_{n2}'$ correspond to $\phi(r)$ and $D_{n2}$ after subtraction of their respective mean values. The phase pattern on SLM, which was used to automatically correct all the aberration during fabrication of lattices, was combination between the phase patterns for correction of the part I and II aberrations. The uniformity and roundness of waveguides across depth fabricated without and with correction of aberration are compared in Fig. S6. The end-on microscope images were taken with a transmission microscope (Zeiss Axioplan 2) at 40× magnification. All the waveguides across depth from 50 to 1000 μm were fabricated with the same writing parameters as those for waveguide lattices fabrications. In Fig. S6, the first column are waveguides fabricated with correction of the part I aberration but not the depth-induced component (part II) and the second column are waveguides fabricated with correction of all the aberrations, from which we can see that the uniformity and roundness of the waveguides are largely preserved across depth though aberration correction. The observed small variation of the waveguides across depths with correction of all the aberrations is thought to be due to residual depth-induced aberration. The importance of the uniformity and maintained roundness of waveguides across depth, which are one of the enabling factors for the success of our experiment, can be understood from fact that functional principle of waveguides lattices are based on evanescent couplings between waveguides that should be as uniform at all transverse directions as possible, although the residual non-uniformity can be compensated to certain degree through design of the lattice (see following section D). As shown in Fig. S7(a), after fabrication, the waveguide roundness and uniformity across depth can be perceived from the end-on microscope image of the waveguide lattices, which was taken with a transmission microscope at 10× and 40× magnification (Zeiss Axioplan 2).

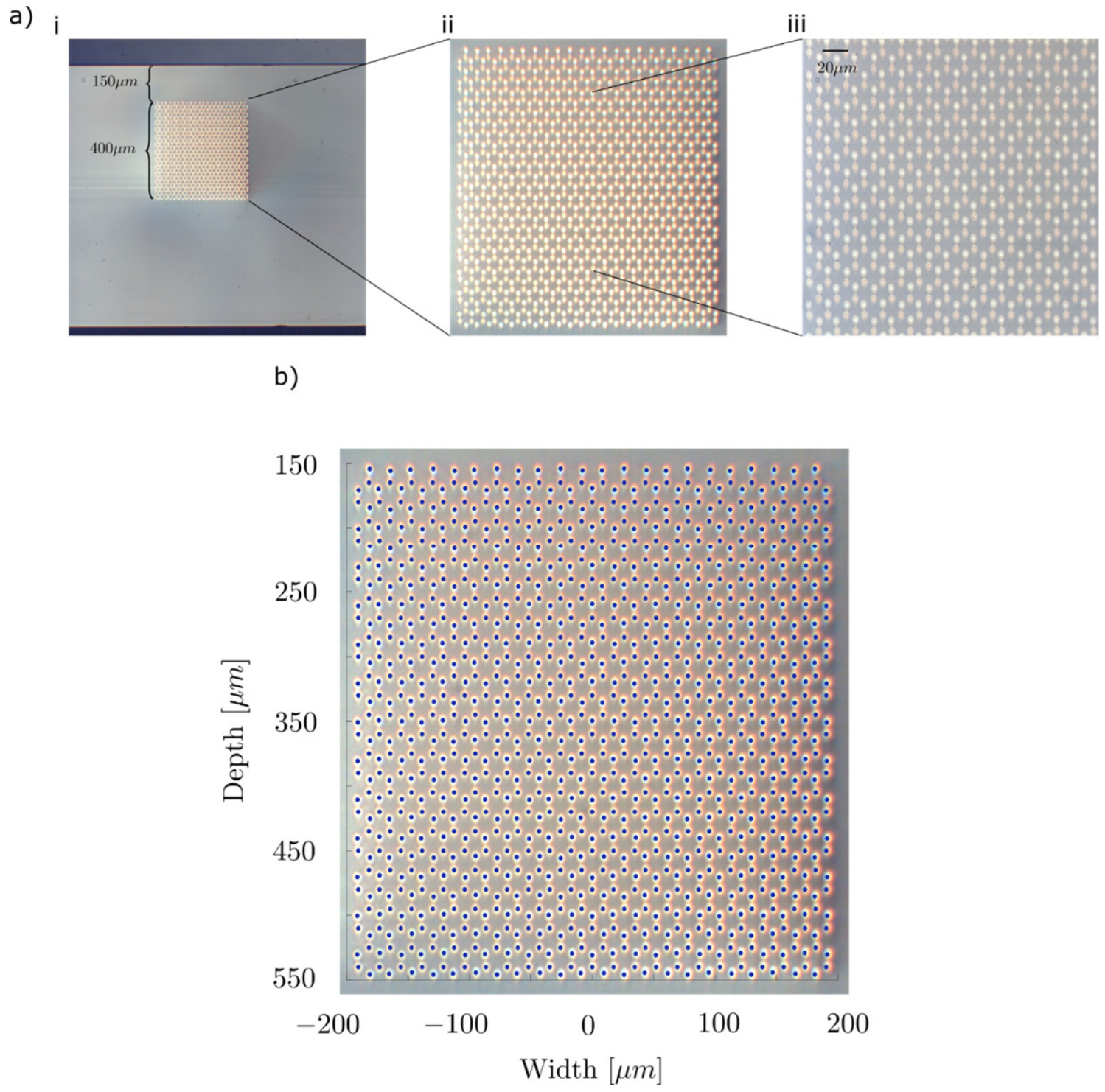

*Fig. S7* a) i) 10× microscope image of the end-facet of the silica chip, the lattice is located at 150 μm from top surface. The chip is 1.1 mm thick. ii) Zoomed in image, showing the entire lattice. iii) 40× microscope image of the centre of the lattice showing the structure of individual wave guides. b) Comparison with desired placement of wave guides. Blue dots mark the designed positions of waveguides.

**D – Characterisation of waveguide coupling strength**

It is important that the waveguide lattice that is written has the correct coupling matrix, which determines the Hamiltonian of the system and thus the eigenmodes of the lattice. To ensure that the fabricated lattice has the correct coupling matrix, the strength of the coupling must be determined and controlled for every position and geometry in the lattice. In this section, we outline the procedure by which we determined the strength of all possible couplings throughout the chip by characterizing well over a thousand beam-splitters in all relevant positions and orientations.

There are two separate effects that must be considered and compensated in our waveguide lattice design to fabricate a chip with the desired coupling matrix:

The first effect is the dependence of coupling strength on the geometry of the coupled waveguides. Because the mode of light in our waveguides is elliptical instead of round, there is a dependence of coupling strength on the relative orientation of the two coupled waveguides. When light in one waveguide couples to another, there is a vector which represents the shortest distance between the two waveguides. We refer to the direction of this vector as the "coupling direction". In a graphene-like lattice there are only two distinct coupling directions. These are the vertical direction and the direction 60 degrees from the vertical. To compensate this effect, we stretch the lattice in one direction to ensure "isotropic" coupling strength between waveguides at different orientations.

The second effect is depth-dependent coupling. Although the waveguides are written with state-of-the-art aberration correction for different depths, the properties of waveguides at different depths are still slightly different. These differences cause a variation in the coupling strength between waveguides. The influence form this effect as measured by the expected near-zero eigenvalue of the zero mode is small enough to be neglected.

To characterise these effects, we fabricated arrays of beam-splitters at different depths with different coupling orientations (vertical, horizontal and diagonal) and separations between two waveguides. For each combination of depth, orientation and separation we wrote multiple directional couplers with varying lengths of the coupling region. This provides us with multiple data points to be fitted to a sine curve for each combination and allows us to determine the coupling strength. The splitting ratios of the directional couplers are measured with the experiment setup described in Section A, in an automatized manner.

The data processing procedure is as follows. The raw data of the splitting ratios for different coupling region lengths is fitted to a sine curve to determine the coupling period (inverse coupling strength). The period for several different separations is then fitted according to an exponential fit (Fig. S8), which shows how the coupling strength varies over separation at a particular depth and orientation.

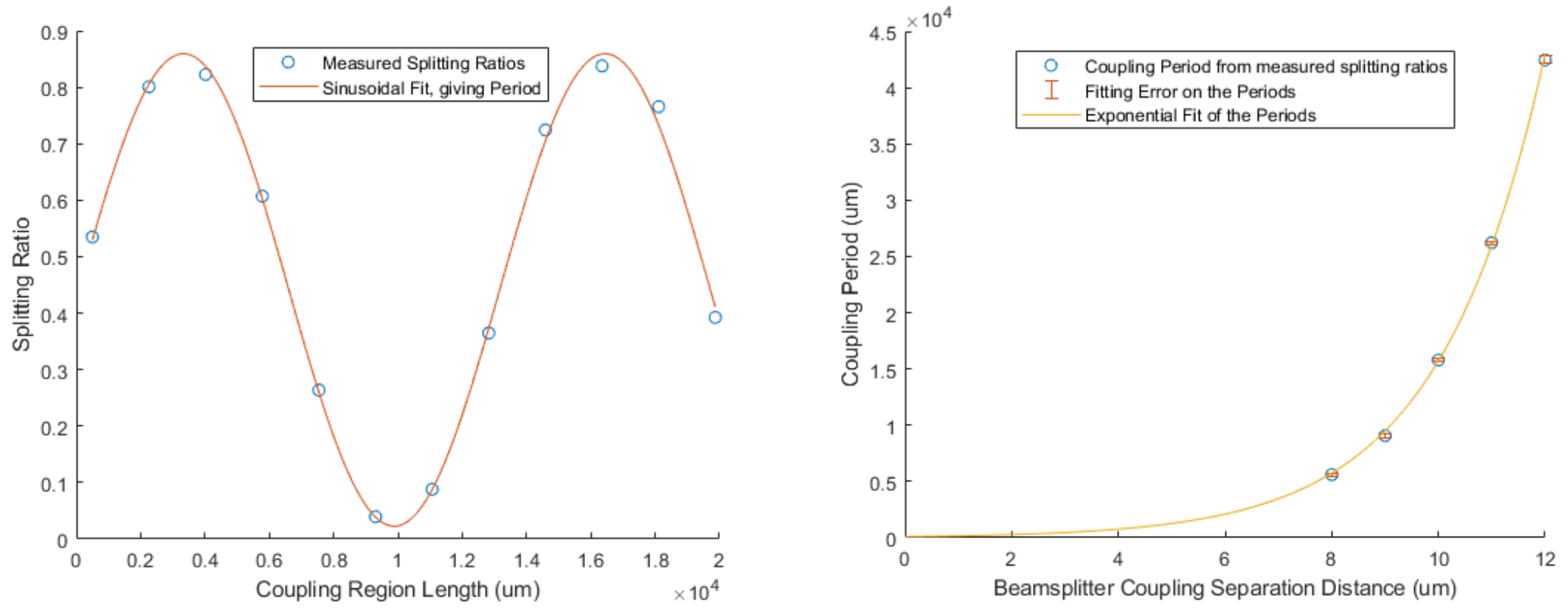

*Fig. S8* Example data from the beam splitter characterisation. This data comes from a depth of 550 µm and is for a diagonal orientation. Left: The splitting ratios for a particular geometry, depth and coupling distance is measured for different coupling region lengths. A sinusoidal fit is then applied to determine the coupling period. Right: The resulting periods for different coupling distances (same geometries and depths) are plotted and fit to an exponential, giving the coupling strength for any distance at this depth in this geometry. In this plot, the data point for 11 µm separation is taken from the fit in the left hand plot.

Once the coupling strength is determined for all waveguide orientations then the waveguide positions can be designed to give the desired coupling matrix. An example of a lattice design with corrected waveguide positions can be seen in Fig. S7(b) in section C.

We numerically calculate the energy spectrum of the real lattice Hamiltonian from the waveguide positions, using experimental data to obtain the coupling strengths. Comparing Fig. S9 a) and b), we can clearly see a band gap opening around zero energy, as the vortex distortion is introduced to the graphene-like waveguide lattice. When there is a bulk zero mode present, this gives rise to a corresponding state at the edge with opposite vorticity $(-1)$, such that the total vorticity of the system vanishes. The topological mode has zero energy and lies together with the edge-states at the centre of the band gap. The energy of the zero-mode is not exactly zero primarily due to the finite size of the lattice. In terms of the oscillation length it is on the order of a few centimetres for a lattice of 1192 sites.

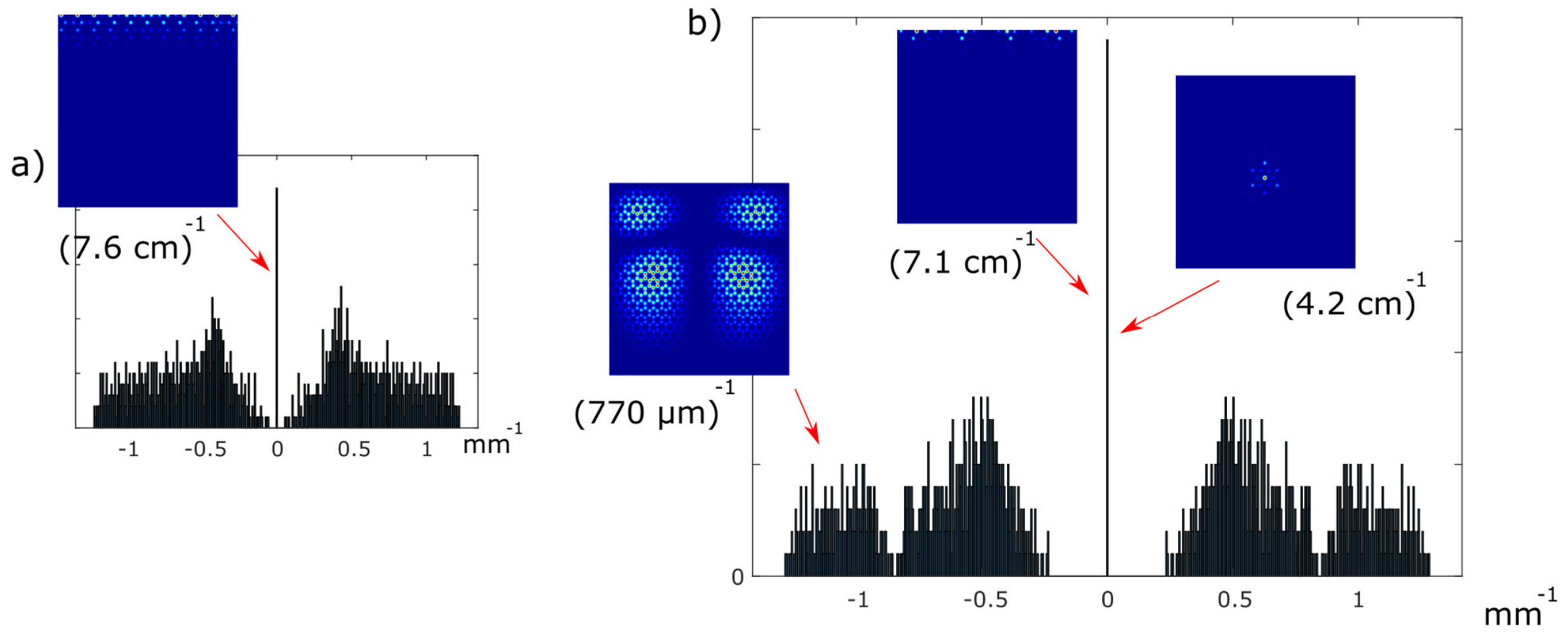

*Fig. S9* a) Numerically calculated density of states of a graphene lattice with 1192 sites. b) Density of states with central vortex distortion. The energy is given in terms of the wavenumber of the mode. Inset images of eigenmodes at different energies are shown. The energy of the zero mode varies on a length scale of several centimetres. In contrast, high energy bulk modes oscillate in the range of a less than a millimetre.